\documentclass[a4paper]{jpconf}
\usepackage{graphicx}
\usepackage[draft]{minted}
\usepackage[sorting=none,giveninits=true,url=false]{biblatex}

\begin{document}
\title{An array-oriented Python interface for FastJet}

\author{Aryan Roy$^1$, Jim Pivarski$^2$, Chad Wells Freer$^3$}

\address{1 Manipal Institute of Technology, Manipal, India}
\address{2 Princeton University, Princeton NJ, USA}
\address{3 Massachusetts Institute of Technology, Boston MA, USA}
\ead{aryanroy5678@gmail.com, pivarski@princeton.edu, chad.freer@cern.ch}

\begin{abstract}
Analysis on HEP data is an iterative process in which the results of one step often inform the next. In an
exploratory analysis, it is common to perform one computation on a collection of events, then view the results
(often with histograms) to decide what to try next. Awkward Array is a Scikit-HEP Python package that
enables data analysis with array-at-a-time operations to implement cuts as slices, combinatorics as composable
functions, etc. However, most C++ HEP libraries, such as FastJet, have an imperative, one-particle-at-a-time
interface, which would be inefficient in Python and goes against the grain of the array-at-a-time logic of
scientific Python. Therefore, we developed fastjet, a pip-installable Python package that provides FastJet C++
binaries, the classic (particle-at-a-time) Python interface, and the new array-oriented interface for use with
Awkward Array.
The new interface streamlines interoperability with scientific Python software beyond HEP, such as machine
learning. In one case, adopting this library along with other array-oriented tools accelerated HEP analysis
code by a factor of 20. It was designed to be easily integrated with libraries in the Scikit-HEP ecosystem,
including Uproot (file I/O), hist (histogramming), Vector (Lorentz vectors), and Coffea (high-level glue). We discuss the design of the fastjet Python library, integrating the classic interface with the array oriented interface and with the Vector library for Lorentz vector operations. The new interface was developed as open source.
\end{abstract}

\section{Introduction}

Jets, an important signal in High Energy Physics (HEP) collisions, are collimated bunches of particles produced by the hadronization of a quark or gluon. They are measured in particle detectors and studied in order to determine the properties of the original quarks. Many jet-finding algorithms were developed independently since 1977, until the FastJet package~\cite{cacciari2012fastjet} was introduced in 2006, containing fast implementations of the most commonly used algorithms. All jet-finding algorithms take a collection of particle 4-momenta (``pseudojets'') and return a collection of jet 4-momenta for each collision event.

\begin{figure}
\begin{minipage}{39pc}
\begin{center} 
\includegraphics[width=0.7\linewidth]{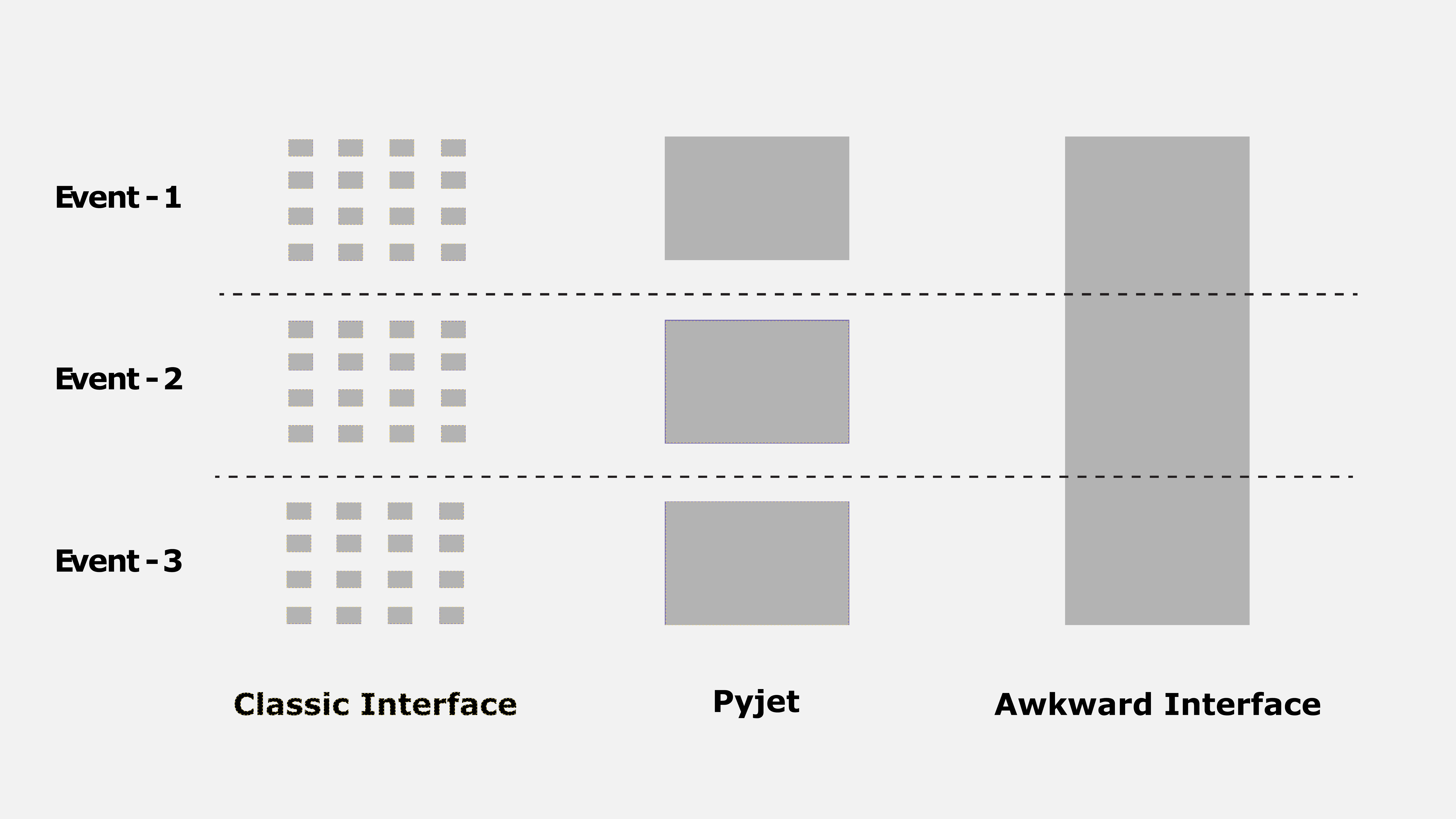}
\caption{\label{fig:vec}Degree of vectorization in FastJet's classic interface (left), in which every particle is represented by a Python object, pyjet (middle), in which every event is a Python object, and Awkward Array (right), in which an entire dataset may be in a single Python object.}
\end{center}
\end{minipage}\hspace{2pc}%
\end{figure}

The mostly-C++ FastJet package includes a Python interface, which addresses each particle of each event as a separate Python object. Loops over Python objects are considerably slower than loops over C++ objects, so many scientific libraries in Python opt for an ``array-oriented'' approach, in which large datasets of numerical arrays are referenced by a single Python object and operations on whole arrays are initiated by individual Python function calls. The pyjet package~\cite{noel_dawe_2021_4446849} in Scikit-HEP~\cite{refId0} has a partially array-oriented interface to FastJet: it refers to each event as a NumPy array of particles (with 4 momentum components each). NumPy cannot describe collections of events, since each event has a different number of input particles and returns different numbers of output jets. Awkward Array~\cite{Pivarski_Awkward_Array_2018}, an extension of NumPy, can represent arrays of varying numbers of particles, and hence it can provide an interface to FastJet that operates on collections of events in a single function call: see Figure~\ref{fig:vec}.

In this work, we present a new Python package for FastJet called fastjet (lowercase) with a fully array-oriented interface through Awkward Array. It will be replacing the pyjet package in Scikit-HEP since pyjet diverged from the original FastJet interface and implements only a subset of its algorithms. This new fastjet package contains the original FastJet bindings, making them available on PyPI (the Python Package Index) so that they can be installed with pip for the first time. From a user's perspective, fastjet {\it is} FastJet in Python, with the original particle-oriented interface, which we call the ``classic'' interface, and the new array-oriented interface in the same namespace. The array-oriented interface uses the same idioms as the classic interface wherever possible, but replaces lists of lists of PseudoJet objects with efficient Awkward Arrays of Vectors for high-volume data.

\section{Features required for the new fastjet package}

The fastjet package was designed with the following constraints:

\begin{itemize}
  \item It must handle deeply nested, variable length lists in an efficient way.
  \item It must perform Lorentz vector calculations on the array elements, including coordinate system transformations.
  \item It must be pip-installable with as many of the FastJet plugins enabled as possible.
\end{itemize}

\section{The Scikit-HEP ecosystem}

The requirements outlined in the previous section can be satisfied by leveraging other packages in the Scikit-HEP ecosystem. The following are directly or indirectly used by fastjet.

\subsection{Awkward Array}

Awkward Array represents and manipulates arrays of data with complex data types, including the lists (and lists of lists) of 4-component momentum records that are the inputs and outputs of jet-finding (including lists of constituents for each jet). Awkward Array can also implement cuts and operate on combinations of particles (``combinatorics''), and it is used for general data analysis by a community of physicists, particularly through the Coffea framework~\cite{lindsey_gray_2021_5762406}. Building a jet-finding interface on Awkward Array gets these benefits for free.

The new fastjet package uses Awkward Arrays in the following form (typically with larger numbers of particles) to represent a collection of events. Non-kinematic fields, like ``charge,'' are not required for jet-finding, but do not need to be excluded from the particle records.

\vspace{-0.4 cm}\begin{equation}\label{eqn:array}
\begin{minipage}{0.9\linewidth}
\begin{minted}{python}
array = ak.Array([
    [
        {"px": 1.2, "py": 3.2, "pz":5.4, "E": 2.5, "charge": 1},
        {"px": 32.2, "py": 64.21, "pz": 543.34, "E": 24.12, "charge": -1},
        {"px": 32.45, "py": 63.21, "pz": 543.14, "E": 24.56, "charge": 1},
    ],
    [],   # empty event
    [
        {"px": 2.95, "py": -0.35, "pz": 0.62, "E": 2.86, "charge": 1},
        {"px": 4.33, "py": 0.53, "pz": 1.47, "E": 3.95, "charge": -1},
        {"px": 0.32, "py": 0.06, "pz": 0.12, "E": 0.21, "charge": 1},
        {"px": 0.32, "py": 0.01579, "pz": 0.01, "E": 0.32, "charge": 1},
    ],
    [
        {"px": 9.74, "py": -0.01, "pz": 0.23, "E": 9.73, "charge": -1},
    ],
])
\end{minted}
\end{minipage}
\end{equation} 

\subsection{Vector}

Vector~\cite{henry_schreiner_2022_5942083} is a library for array-oriented operations on 2D, 3D and Lorentz vectors. The new fastjet library uses Vector to perform coordinate transformations, calculate angles between vectors, $\Delta R$, etc. Since our arrays of particles bypass FastJet's PseudoJet class, Vector is needed to replace the functionality that PseudoJet's methods would ordinarily be used for.

\section{The particle-oriented interface} 

 Clustering many events in one function call reduces the amount of code users need to write, thereby improving the ergonomics of the analysis workflow. A typical analysis using the classic FastJet interface with the \mintinline{python}{array} variable defined in Equation~(\ref{eqn:array}) would look like this:
 
\vspace{-0.4 cm}\begin{equation}\label{eqn:classic}
\begin{minipage}{0.9\linewidth}
\begin{minted}{python}
jetdef = fastjet.JetDefinition(fastjet.antikt_algorithm, 0.6)
for original_event in array:
    fastjet_event = []
    for p in original_event:
        fastjet_event.append(
            fastjet.PseudoJet(p["px"], p["py"], p["pz"], p["E"])
        )
    cluster = fastjet.ClusterSequence(fastjet_event, jetdef)
    jets = cluster.inclusive_jets()
\end{minted}
\end{minipage}
\end{equation}

\noindent But the same analysis in the array-oriented interface looks like this:

\vspace{-0.4 cm}\begin{equation}\label{eqn:array-oriented}
\begin{minipage}{0.9\linewidth}
\begin{minted}{python}
jetdef = fastjet.JetDefinition(fastjet.antikt_algorithm, 0.6)
cluster = fastjet.ClusterSequence(array, jetdef)
jets = cluster.inclusive_jets()
\end{minted}
\end{minipage}
\end{equation}

\noindent The classic interface requires explicit loops over particle data, while the array-oriented interface achieves the same result with considerably less code. Since the loops are now in the C++ code invoked by the ClusterSequence constructor, it is also much faster.

\begin{figure}
\begin{minipage}{39pc}
\begin{center}
\includegraphics[width=0.6\linewidth]{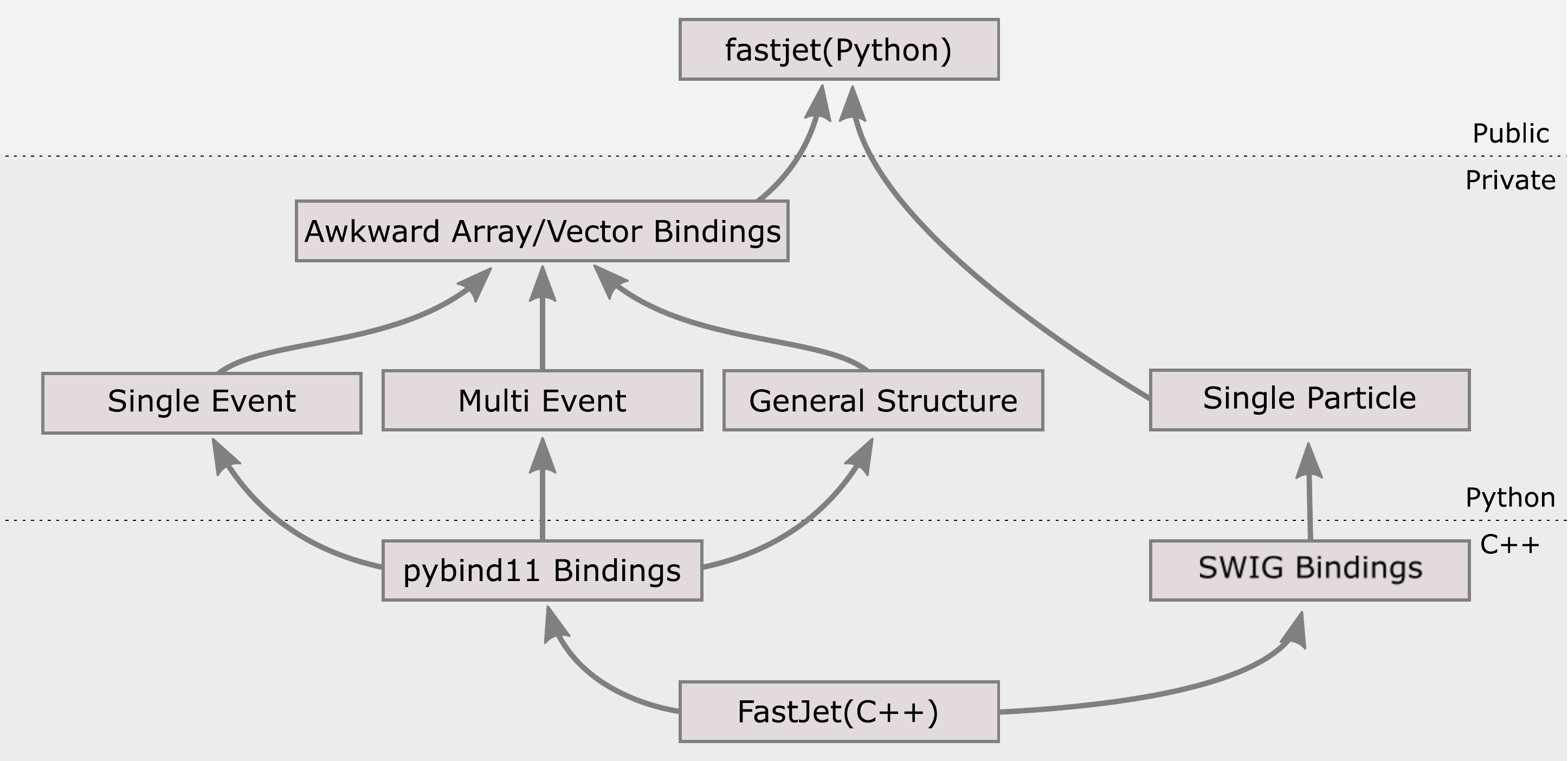}
\caption{\label{fig:design}The internal structure of fastjet; each box represents a distinct part of the code and arrows represent dependencies between them.}
\end{center}
\end{minipage}\hspace{2pc}%
\end{figure}

\section{The internal structure}

The array-oriented interface is implemented with new Python-C++ bindings, written with pybind11~\cite{pybind11}. The classic interface is implemented with SWIG~\cite{10.5555/1267498.1267513}, another (much older) tool for interfacing Python and C++. Both of these bindings are included in the fastjet package so that users can access either one. This two-bindings architecture is depicted in Figure~\ref{fig:design}.


At its core, FastJet is a C++ library with a C++ interface. The Python bindings that are included with it are a direct mirror of the C++ interface: every C++ class and method has a corresponding Python class and method. This allows for easy porting between C++ and Python and requires only one suite of documentation, though the resulting Python code is slower and more verbose. Since the fastjet package includes FastJet shared object libraries, it also includes these original Python bindings and makes them visible in the public namespace.


The array-oriented bindings directly connect new Python functions to C++, without going through the original bindings.

Awkward Arrays can have a variety of structures, and jet-finding should be possible for any structure that at least has a collection of records with 4-momentum attributes (in any coordinate system Vector supports). The most basic case is ``array of lists (events) of particle records.'' Other cases can be implemented in terms of this one, such as an ``array of particle records (representing one event),'' which can be viewed as a basic case of length 1. Also, general data structures that eventually contain ``lists (events) of particle records,'' possibly within other records and lists, can be implemented from the basic case while preserving the structures it is nested within. A tree containing arrays of the right form is spliced to provide the right output.

Jet-finding only needs 4 kinematic variables, $p_x$, $p_y$, $p_z$, $E$ in Cartesian coordinates. Any other fields are allowed and passed through fastjet functions that return the original constituents, without interpretation. Including extra fields, such as the ``charge'' in Equation~\ref{eqn:array}, is equivalent to setting up user data in FastJet's PseudoJet class, though the latter relies on \mintinline{c++}{void*} pointer handling. The array-oriented view provides this necessary feature in a natural, less rigid way.

To adhere to the original interface as much as possible, arguments that do not need to be arrays accept the same argument types as the corresponding classic functions, such as JetDefinition in Equation~\ref{eqn:array-oriented}. SWIG objects can supply their C++ pointer values through Python and pybind11 can call Python, so code written with pybind11 can directly access the C++ objects that SWIG wraps: the two-bindings architecture does not prevent interoperability.

\section{Case Study}

The new fastjet package is an essential part of a search for soft unclustered energy patterns (SUEPs) in the CMS detector at the LHC. This search is characterized by a large number of low-pT tracks~\cite{Knapen:2016hky}. As such, non-standard jet-finding features prominently in this analysis. The analysis was originally performed using pyjet in the NanoAODTools framework~\cite{Rizzi:2019rsi}, but was recently rewritten with Awkward Array for data manipulation, Vector for Lorentz kinematics, Coffea for scale-out, and fastjet for jet-clustering. The new workflow can analyze 50~thousand events in under a minute, as opposed to 17~minutes in the original workflow, while producing exactly the same results. The throughput also increased from 40~Hz to 1000~Hz.

Used in conjunction with the other Scikit-HEP packages, fastjet simplifies the user experience by not having to switch from an array-oriented perspective to a particle-oriented perspective just for jet-finding, and this encourages the use of higher-performance tools.

\section{Conclusion}

In this paper, we presented the design choices and performance indicators of a new array-oriented interface for jet-finding with FastJet. The new package is pip-installable and seemlessly integrates with other Scikit-HEP packages such as Awkward Array and Vector. The array-oriented design results in improved performance when dealing with large sets of events, such as the factor of 20 observed in the search for SUEPs at the LHC. Representing kinematic data in Awkward record arrays, which are less rigid than C++ PseudoJet instances, allows user data to be attached in a natural way.

This package brings the new paradigm of array-oriented analysis to jet clustering, allowing users to think about these workflows in a new way. Making temporary arrays of jets clustered with different settings becomes very easy in an array-based interface, and fastjet's connections to other libraries in the Scikit-HEP ecosystem opens the door to a wider set of possibilities.

\section{Acknowledgements}

We would like to thank Matteo Cacciari, Gavin Salam, Gregory Soyez, Salvatore Rappoccio, Patrick Komiske, and Eduardo Rodrigues for helpful discussions. This work was supported by the National Science Foundation under Cooperative Agreement OAC-1836650 (IRIS-HEP).

\printbibliography

\end{document}